\documentclass[apj]{emulateapj}

\newcommand {\Lya}    {Ly$\alpha$}   

\newcommand {\Ha}     {H$\alpha$}

\newcommand {\HI}     {\ion{H}{1}}   
\newcommand {\NV}     {\ion{N}{5}}
\newcommand {\CII}    {\ion{C}{2}}
\newcommand {\SiIII}  {\ion{Si}{3}}
\newcommand {\SiIV}   {\ion{Si}{4}}
\newcommand {\OI}     {\ion{O}{1}}
\newcommand {\SiII}   {\ion{Si}{2}}
\newcommand {\NaI}    {\ion{Na}{1}}
\newcommand {\CaII}   {\ion{Ca}{2}}
\newcommand {\MgI}    {\ion{Mg}{1}}
\newcommand {\MgII}   {\ion{Mg}{2}}
\newcommand {\MnII}   {\ion{Mn}{2}}
\newcommand {\FeII}   {\ion{Fe}{2}}
\newcommand {\FeXXI}  {[\ion{Fe}{21}]}

\newcommand {\kms}    {km~s$^{-1}$}
\newcommand {\NHI}    {$N_{\rm HI}$}

\newcommand {\tnma}{\tablenotemark{a}}
\newcommand {\tnmb}{\tablenotemark{b}}

\newcommand {\nd}  {\nodata}

\newcommand {\etal}  {et~al.} 
  
\newcommand {\flux}  {$\rm erg~cm^{-2}~s^{-1}~\AA^{-1}$}

\shorttitle{FR\,1 Galaxies in the Far-UV}
\shortauthors{Danforth et al.}

\begin{document}

\title{Far-UV Emission Properties of FR\,1 Radio Galaxies
\footnote{Based on observations made with the NASA/ESA {\it Hubble Space Telescope}, obtained from the data archive at the Space Telescope Science Institute. STScI is operated by the Association of Universities for Research in Astronomy, Inc. under NASA contract NAS5-26555.}}

\author{Charles W. Danforth\altaffilmark{1}, John T. Stocke\altaffilmark{1}, Kevin France\altaffilmark{1,2}, Mitchell C. Begelman\altaffilmark{3}}
\affil{Department of Astrophysical \& Planetary Sciences, University of Colorado, 391-UCB, Boulder, CO, 80309, USA; danforth@colorado.edu}
\and
\author{Eric Perlman}
\affil{Department of Physics \& Space Sciences, Florida Institute of Technology, 150 W. University Blvd., Melbourne, FL, 32901}
\altaffiltext{1}{Center for Astrophysics and Space Astronomy, University of Colorado, 389-UCB, Boulder, CO, USA 80309}
\altaffiltext{2}{Laboratory for Atmospheric and Space Physics, University of Colorado, 600-UCB, Boulder, CO 80309}
\altaffiltext{3}{JILA \& Department of Astrophysical \& Planetary Sciences, University of Colorado, 391-UCB, Boulder, CO, USA 80309}

\begin{abstract}
The power mechanism and accretion geometry for low-power FR\,1 radio
galaxies is poorly understood in comparison to Seyfert galaxies and
QSOs.  In this paper we use the diagnostic power of the \Lya\
recombination line observed using the Cosmic Origins Spectrograph
(COS) aboard the Hubble Space Telescope (HST) to investigate the
accretion flows in three well-known, nearby FR\,1s: M\,87, NGC\,4696,
and Hydra\,A.  The \Lya\ emission line's luminosity, velocity
structure and the limited knowledge of its spatial extent provided by
COS are used to assess conditions within a few parsecs of the
super-massive black hole (SMBH) in these radio-mode AGN.  We observe
strong \Lya\ emission in all three objects with similar total
luminosity to that seen in BL\,Lacertae objects.  M\,87 shows a
complicated emission line profile in \Lya\ which varies spatially
across the COS aperture and possibly temporally over several epochs of
observation. In both NGC\,4696 and M\,87, the \Lya\ luminosities
$\sim10^{40}$ ergs~s$^{-1}$ are closely consistent with the observed
strength of the ionizing continuum in Case B recombination theory and
with the assumption of near unity covering factor. It is possible that
the \Lya\ emitting clouds are ionized largely by beamed radiation
associated with the jets.  Long-slit UV spectroscopy can be used to
test this hypothesis.  Hydra\,A and the several BL\,Lac objects
studied in this and previous papers have \Lya\ luminosities larger
than M\,87 but their extrapolated, non-thermal continua are so
luminous that they over-predict the observed strength of \Lya, a clear
indicator of relativistic beaming in our direction.  Given their
substantial space density ($\sim4\times10^{-3}$ Mpc$^{-3}$) the
unbeamed Lyman continuum radiation of FR\,1s may make a substantial
minority contribution ($\sim10$\%) to the local UV background if all
FR\,1s are similar to M\,87 in ionizing flux level.
\end{abstract}



\section{Introduction} 

Low power Fanaroff-Riley (1974) class 1 (FR\,1) radio galaxies
\citep[$P_{20cm}\leq 10^{24-25}~\rm W\,Hz^{-1}$; ][]{LedlowOwen96} are
among the most numerous AGN, comparable in space density to the
better-known (and far better-understood) Seyfert galaxies
\citep{MauchSadler07,Owen96,HoUlvestad01}.  M\,87 with its famous
synchrotron jet is the FR\,1 class prototype with other local examples
including Centaurus\,A, NGC\,4696 (the brightest galaxy in the
Centaurus Cluster), and Hydra\,A.  But where Seyfert galaxies exhibit
luminous accretion disk emission and broad-line-region (BLR) emission
lines, these two emission mechanisms are mostly or entirely absent in
FR\,1s, making their power source more mysterious.  Adding to this
mystery is the presence of FR\,1 low-power sources in most brightest
cluster galaxies (BCGs), including the morphologically-distinct ``cD''
galaxies \citep{Lauer14}.  For example, 70\% of BCGs in clusters with
a ``cool X-ray core'' have FR\,1 radio sources with $P_{20cm}\geq
10^{20}$ W Hz$^{-1}$ \citep{Burns90}; \citep[see also][]{LedlowOwen96,
Stocke99, Perlman04, Branchesi06, Blanton11}; i.e., FR\,1s are
commonly found in clusters and rich groups of galaxies and often, but
not exclusively in the BCG \citep{LedlowOwen95, LedlowOwen96}.

These BCGs are the most luminous, most massive galaxies in the
Universe.  Given the observed correlation between the supermassive
black hole (SMBH) mass and the bulge mass or bulge velocity dispersion
\citep{KormendyHo13}, these galaxies should also possess the most
massive SMBHs.  Indeed, local FR\,1 galaxies host some of the most
massive SMBHs known: e.g., M\,87 \citep[$3.5^{+0.9}_{-0.7}\times10^9
~\rm M_\sun$;][]{Walsh13} and two rich cluster BCGs \citep[$1.3$ and
$3.7\times10^9\rm M_\sun$; listed in ][]{KormendyHo13}.  However these
most massive SMBHs power very low-luminosity AGN in comparison to many
found in lower-mass, lower-luminosity ellipticals and early-type
spirals.  Apparently these sources either have little fuel available
or do not reprocess gravitational energy as effectively into X-rays,
relativistic particles, and radio emission as many AGN in
less-luminous galaxies.  It is possible that these sources do extract
enormous amounts of energy from these most-massive black holes but
that most of this energy is converted into the kinetic energy of a
bulk outflow whose effect is visible as giant cavities in the
surrounding X-ray emitting gas \citep[e.g.,][]{McNamaraNulsen06,
Fabian11}.  Alternatively, it is possible that for some reason the
sub-parsec structures near the SMBH prevent these sources from being
as luminous as AGN in spirals (i.e., Seyferts and QSOs) and even those
in lower luminosity ellipticals (i.e., FR\,2 radio galaxies and
quasars).

Theoretical modeling of FR\,1s \citep[sometimes called the AGN ``radio
mode'';][]{Croton06} rely on a ``radiatively inefficient accretion
flow'' \citep[RIAF, or its several variants:][]{NarayanYi94,
NarayanYi95, BlandfordBegelman99,QuataertGruzinov00,HawleyBalbus02},
but its parameters are much more poorly quantified than for luminous
accretion disk AGN.  For example, \citet[][A06 hereafter]{Allen06}
found a strong correlation between the Bondi accretion rate (the hot
gas mass available at or near the Bondi radius) and the observed jet
power in FR\,1s.  While A06 use this result to suggest that Bondi
accretion is a good model for an RIAF, they also derive a very high
efficiency (2\%) for converting the energy of the gas accretion into
relativistic jet power, requiring a very efficient accretion flow from
the Bondi radius ($\sim100-200$ parsecs) all the way into the SMBH.
Detailed models of this process \citep[e.g.,][]{NarayanYi95,
BlandfordBegelman99} infer that such efficient accretion is quite
unlikely and that in many circumstances a thin accretion disk and an
outflowing wind can be formed as in the ``ADIOS'' (advection-dominated
inflow-outflow solution) model of \citet{BlandfordBegelman99}.  For
example, \citet{BegelmanCelotti04} argue for the ``ADIOS'' scenario in
which the presence of an ``outer accretion disk'' and an outflowing
disk wind will reduce the accretion flow by two orders of magnitude.

Observationally, deep {\it Chandra} imaging of the FR\,1 prototype
M\,87 derived an accretion rate comparable to the expected Bondi rate
of 0.1 $M_{\odot}$ yr$^{-1}$ \citep{DiMatteo03} outside the Bondi
radius ($\sim100-200$ pc).  However, more recent X-ray imaging
spectroscopy work by \citet{Russell15} finds an accretion flow greatly
reduced below the Bondi rate inside the Bondi radius of M\,87.

Detailed fitting of the spectral energy distribution (SED) within
$\sim$ 30 pc of the nucleus of M87 by \citet{Prieto16} finds an SED
dominated by jet emission throughout the electromagnetic spectrum
excepting a small contribution from a cool accretion disk longward of
$1\mu$m. This corroborates earlier work by \citet{Perlman01,
Perlman07}.  Based on the weak, cool disk detected in the IR,
\citet{Prieto16} derive a very low accretion rate of $<6\times
10^{-5}~M_{\odot}$ yr$^{-1}$ from the small, cool accretion disk they
derive to be present from the SED. This same broadband SED was also
used by \citet{Broderick15} to place similar constraints on the
accretion rate and on the presence of an event horizon.
\citet{Prieto16} also derive a similarly low accretion rate from the
previous HST imaging data on the \Ha\ emission line in the ionized gas
disk discovered by \citet{Ford94}. The absence of significant Faraday
rotation for the M\,87 nuclear source at sub-mm wavelengths is also
consistent with a very low accretion rate \citep{Kuo14}.  Clearly
these observational and theoretical results cause problems for the
simple Bondi rate model of A06, suggesting that a more complex nuclear
geometry than a spherically symmetric, hot gaseous atmosphere is
necessary in M\,87 and other similar sources.

 \begin{deluxetable*}{lllcccl}  
 \tabletypesize{\footnotesize}  
 \tablecolumns{7}   
 \tablewidth{0pt}   
 \tablecaption{Summary of Observations}  
 \tablehead{
	\colhead{Galaxy} & 
	\colhead{RA (J2000) Dec} & 
	\colhead{$z_{\rm AGN}$} &
	\colhead{Galactic} & 
	\colhead{HST} &
	\colhead{$t_{\rm exp}$} & 
	\colhead{Obs.}  
	\\ 
	\colhead{} & 
	\colhead{} &
	\colhead{} & 
	\colhead{$E(B-V)$} &
	\colhead{Program} &
	\colhead{(ksec)} &
	\colhead{Date} 
	}
\startdata 
  \cutinhead{FR\,1 Galaxies}
    M\,87             & 12 30 49.4  $+$12 23 27 & 0.0044 & 0.020 & 13489 & 4.4 & 2014, Mar 3\\ 
    NGC\,4696         & 12 48 49.2  $-$41 18 39 & 0.00987& 0.098 & 13489 & 5.2 & 2013, Dec 20\\
    Hydra\,A          & 09 18 05.7  $-$12 05 44 & 0.05488& 0.036 & 13489 & 4.7 & 2013, Nov 6\\
  \cutinhead{BL\,Lac Objects}						 
    1ES\,1028$+$511   & 10 31 18.5  $+$50 53 36 & 0.3604 & 0.012 & 12025 & 22,33& 2011, May 1\\
    PMN\,J1103$-$2329 & 11 03 37.6  $-$23 29 30 & 0.186  & 0.078 & 12025 & 18,13& 2011, Jul 5 \\
  \enddata  
\end{deluxetable*}

Because an accretion disk and wind can create the physical conditions
necessary for a broad-line-region (BLR) at sub-parsec scales, \Lya\
and other UV emission lines can provide a method for probing the
accretion region in FR\,1s much closer to the black hole than can
X-ray continuum observations (i.e., 2--30~pc in nearby FR\,1s).
Observations using the full spatial resolution of HST (0\farcs025 in
the UV) are capable of resolving line and continuum emission at the
level of 2~pc in M\,87, 4.5~pc in NGC\,4696 and 30~pc in Hydra\,A.
Well-sampled optical imaging of M\,87 in \Ha\ \citep{Ford94} and
imaging spectroscopy in [\ion{O}{2}] \citep{Macchetto97} and \Ha\
\citep{Walsh13} with the {\it Space Telescope Imaging Spectrograph}
(STIS) yield detections of this disk to within $\sim$ 6 pc of the
SMBH. UV spectroscopy can do a bit better than this so the detection
of \Lya\ through the 0."26 FOS aperture by \citet{Sankrit99} provides
a probe of ionized gas close to the nucleus as well ($\sim20$ pc). The
detection of \Lya\ emission in M\,87 and other FR\,1s verifies the
presence of cooler gas in the RIAF and its observed line-width and
shape can determine some specifics about the accretion; e.g., the
minimum mass that has ``dropped out'' of a hot accretion flow.  Also,
the mere presence of small amounts of cool gas inside the Bondi radius
supports models of a much more inefficient RIAF than originally
suggested \citep{BlandfordBegelman99, BegelmanCelotti04,
SikoraBegelman13}, consistent with the new X-ray analysis of
\citet{Russell15} and the SED analysis of \citet{Prieto16}.  Weak
emission line wings also are expected in some models
\citep{SikoraBegelman13} as well as weak, highly-ionized absorption
suggestive of an outflowing wind (an ``ADIOS'').

Weak optical emission lines have been observed in some FR\,1s
\citep[e.g.,][]{LedlowOwen95,Ho97,Ho01}, but observed optical
line-widths are usually rather narrow ($\leq300$ \kms) and the $\rm
[N\,II]/H\alpha$ ratios are indicative of LINER emission \citep{Ho01}.
There is no indication in most cases that much of this emission line
gas is participating in the accretion process.  Given the dramatic
spatial extent of \Ha\ emission in some of the nearest FR\,1s,
including M\,87 \citep{Ford94, Ford97} and NGC\,4696
\citep{Canning11}, most of the observed narrow emission line gas is
not nuclear at all.  However, some few BL\,Lac objects have been
observed to possess broad \Ha\ lines, including OJ\,287
\citep{SitkoJunkkarinen85}, PKS\,0521$-$365 \citep{Ulrich81} and even
BL\,Lac itself \citep{Vermeulen95}, which is undoubtedly nuclear.

Recently, using the {\it Cosmic Origins Spectrograph} (COS) on the
Hubble Space Telescope (HST), we discovered weak, and relatively
narrow \Lya\ emission lines in three low-$z$ BL\,Lac objects (Stocke,
Danforth \& Perlman 2011; Paper~1 hereafter).  The COS spectra provide
\Lya\ emission line luminosities ($10^{40-41}~{\rm erg ~s^{-1}}$),
line widths ($FWHM=200-1000$ \kms) and simultaneous beamed continuum
luminosities.  These observed \Lya\ line luminosities are orders of
magnitude less than those seen in Seyfert Galaxies and are well below
the line luminosities required in an ionization-bounded BLR based on
the observed non-thermal continuum luminosities in these sources.  But
we know that the continuum is beamed in BL\,Lacs and an uncertain
beaming factor needs to be applied to estimate the luminosity of the
unbeamed continuum to determine whether the cool \Lya-emitting gas is
ionization-bounded or density-bounded.  If the gas surrounding the
SMBH in FR\,1 AGN is density-bounded (the prediction of RIAF models),
the \Lya\ line luminosities provide an order-of-magnitude estimate of
the amount of cool gas in the BLR.

In this paper we use far-UV (FUV) spectroscopic observations from
HST/COS of three nearby FR\,1s: M\,87, NGC\,4696 and Hydra\,A and two
unpublished BL\,Lac objects to investigate the relationship between
FR\,1s and BL\,Lac objects and to investigate the \Lya\ and continuum
luminosities in unbeamed radio galaxies.  \Lya\ emission lines
consistent with being circum-nuclear in origin are seen in all three
FR\,1s.  Therefore, these emission lines and the unbeamed continua
underlying them can be used to probe accretion physics.  In Section~2
we present and describe the COS spectroscopy and data modeling
techniques for these three FR\,1s as well as for two
as-yet-unpublished BL\,Lac objects from observing time provided to the
COS Science Team.  In Section~3 we describe our findings and derive
basic physical parameters from emission models. In Section~4 we
discuss the implications of these observations for physical models of
the SMBH and circum-nuclear regions of these objects.  Finally, in
Section~5 we summarize our main results and discuss their importance
for AGN physics and cosmology.


\section{Observations and Data Analysis}
 
Three FR\,1 galaxies (Table~1) were observed as part of HST Guest
Observer (GO) program 13489 (PI: Stocke).  Each was observed for a
nominal two-orbit visit using the COS/G130M grating.  Four exposures
at different central wavelengths give continuous spectral coverage
over the entire G130M range ($1140\la\lambda\la1465$) at a resolution
of $R\approx17,000$ ($\Delta v\sim15$ \kms).

The two BL\,Lac objects in Table~1 were observed as part of the COS
Guaranteed Time Observations (GTO) programs (PI: Green).  Like the
previous three BL\,Lacs for which \Lya\ emission was detected by
Paper~1, these were observed with both the COS/G130M and G160M
gratings giving continuous spectral coverage over $1140<\lambda<1795$
\AA\ at a resolution of $R\approx17,000$.  Intervening absorption line
measurements for these two BL\,Lacs were published as part of the
\citet{Danforth16} compilation, but AGN continuum and emission-line
properties are new to this work.

The data were retrieved from the Mikulski Archive for Space Telescopes
(MAST) and reduced via the methods described in \citet{Danforth10,
Danforth16}.  Briefly, each exposure was binned by three native pixels
or $\sim40\%$ of a point-source resolution element to increase the
signal-to-noise before coaddition.  The exposures were coaligned in
wavelength by cross-correlating strong Galactic absorption features in
each exposure and interpolated onto a common wavelength vector and
then combined using an exposure-time-weighted algorithm.  More details
of the COS spectral reduction process can be found in
\citet{Danforth16} and \citet{Keeney12}.

Next, the Galactic \Lya\ profile was fitted to determine the Milky Way
\NHI\ which was in turn used to remove foreground reddening
(calculated $E(B-V)$ values are given in Table~1).  In all three FR\,1
cases, the foreground reddening is small.  No internal source
reddening was assumed for either the FR\,1s or the BL\,Lacs.

\begin{figure*}
   \epsscale{.95}\plotone{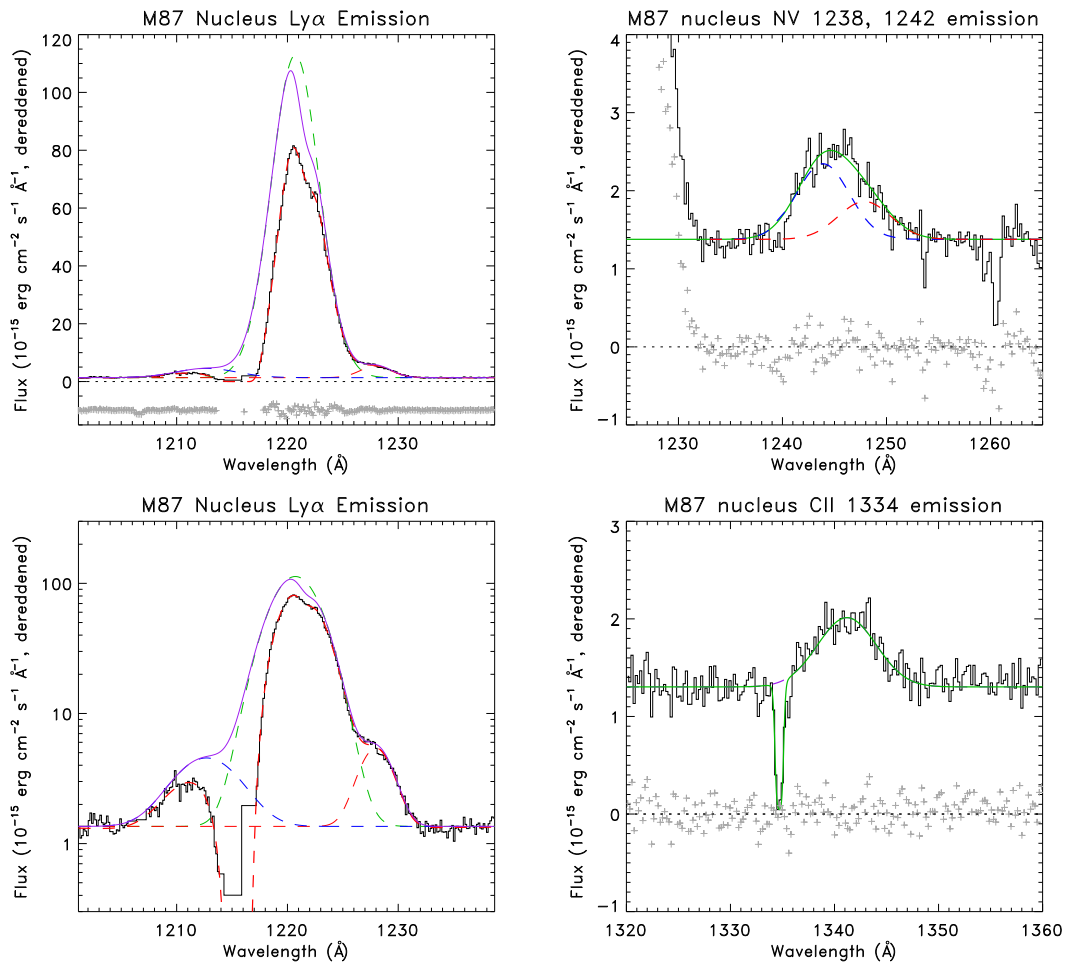}
   \caption{Emission line profiles and model fits for M\,87.  Strong
   \Lya\ emission is seen in the M\,87 nucleus (upper left).  A closer
   look (lower left) shows two additional weak emission features
   ($\sim2-4\%$ of the flux) as well.  The emission profile is
   modified by the Galactic Damped \Lya\ absorption and a broad,
   shallow absorption component as described in the text.  Clear
   emission profiles from the N\,V doublet (upper right) and C\,II
   line (lower right) are also seen.  The obvious absorption at 1334
   \AA\ (lower right panel) is Galactic C\,II.  See text for full
   details of all fitted parameters.}
\label{fig:m87lines}
\end{figure*}

\section{Results} 

All three FR\,1 galaxies show prominent \Lya\ emission at very close
to the systemic redshifts.  Both M\,87 and Hydra\,A show weak FUV
continuum emission while the continuum level in NGC\,4696 is barely
detectable in these data.  In M\,87 the \Lya\ emission line has
extensive line wings to both blue- and red-shifted relative
velocities.  In addition, M\,87 shows weak emission at $\sim1240$ \AA,
$\sim 1340$ \AA, and $\sim1410$~\AA\ which we identify with \NV, \CII,
and \SiIV$+$\ion{O}{4}], respectively.  No metal ion emission was seen
in either of the other two FR\,1s.  Both new BL\,Lac objects
(1ES\,1028$+$511 and PMN\,J1103$-$2329) show smooth power-law continua
with weak \Lya\ emission at approximately the systemic redshifts.  We
discuss the detailed measurements of each source below.

\begin{deluxetable*}{lcccl}  
\tabletypesize{\footnotesize}  
\tablecolumns{5}   
\tablewidth{0pt}   
\tablecaption{M\,87 Emission Line fits}  
\tablehead{\colhead{Quantity}   &             
\colhead{Blue comp.}&             
\colhead{Main comp. }& 	    
\colhead{Red comp.}&
\colhead{Unit}    }  	
\startdata 
\cutinhead{H\,I \Lya\ 1215 \AA}
 	 centroid  &$ 1211.93\pm0.47$ &$1220.79\pm0.05 $&$ 1228.00\pm0.07$& \AA \\
 	           &$-923\pm116$      &$   1263\pm12   $&$   3040\pm17   $& \kms \\
 	 FWHM      &$    6.43\pm0.68$ &$   5.04\pm0.08 $&$   3.30\pm0.15 $& \AA \\
  	           &$    1600\pm170 $ &$   1241\pm18   $&$    813\pm37   $& \kms \\
 $I_{\rm Ly\alpha}$&$      22\pm5   $ &$ 597\pm16$&$  14.0\pm0.7 $& $10^{-15}\rm~erg~cm^{-2}~s^{-1}$ \\
 $L_{\rm Ly\alpha}$&$    0.18\pm0.15$ &$   5.06\pm0.35$\tnma&$0.12\pm0.02$& $10^{39}\rm~erg~s^{-1}$ \\
 	 continuum &  \nd             &$   1.20\pm0.03 $&   \nd           & $10^{-15}\rm~erg~cm^{-2}~s^{-1}~\AA^{-1}$ \\
 \cutinhead{N\,V 1238, 1242 \AA} 
	 centroid  &\nd &$    1226\pm30  $&\nd & \kms \\
 	 FWHM      &\nd &$     5.5\pm0.8 $&\nd &  \AA \\
 	           &\nd &$    1443\pm100 $&\nd & \kms \\
  $I_{\rm NV 1238}$&\nd &$     4.9\pm0.9 $&\nd & $10^{-15}\rm~erg~cm^{-2}~s^{-1}$ \\
  $I_{\rm NV 1242}$&\nd &$     3.7\pm0.7 $&\nd & $10^{-15}\rm~erg~cm^{-2}~s^{-1}$ \\
   doublet ratio   &\nd &$     1.3\pm0.3 $&\nd & \\
         continuum &\nd &$    1.27\pm0.02$&\nd & $10^{-15}\rm~erg~cm^{-2}~s^{-1}~\AA^{-1}$\\
 \cutinhead{C\,II 1334 \AA}        	        
 	 centroid  &\nd &$    1500\pm40 $ &\nd & \kms \\
 	 FWHM      &\nd &$    6.76\pm0.88$&\nd &  \AA \\
  	           &\nd &$    1500\pm100$ &\nd & \kms \\
 $I_{\rm CII}$     &\nd &$    4.59\pm0.59$&\nd & $10^{-15}\rm~erg~cm^{-2}~s^{-1}$ \\
         continuum &\nd &$    1.20\pm0.02$&\nd & $10^{-15}\rm~erg~cm^{-2}~s^{-1}~\AA^{-1}$
  \enddata 
  \tablenotetext{a}{assuming $D=16.7$ Mpc}  
\end{deluxetable*}


\subsection{M\,87} 

The nucleus of the FR\,1 prototype M\,87 with its famous
optical/radio/X-ray jet has been observed intensely over the last two
decades with HST including previous UV and optical spectra obtained
with the Faint Object Spectrograph (FOS) by \citet{Tsvetanov98,
Tsvetanov99} and \citet{Sankrit99}.  These FOS spectra of the nucleus
cover approximately the same wavelength range as the COS/G130M data
including the \Lya\ emission in M\,87 at 1220.8~\AA.  In both the two
\citet{Sankrit99} observations and our own, the intrinsic flux and
line shape of \Lya\ are difficult to determine due to the presence of
the Galactic damped \Lya\ absorption (DLA) as well as probable
absorption in M\,87.  Regardless of the corrections which must be made
due to the Galactic DLA and other absorbers in M\,87,
\citet{Sankrit99} find that the \Lya\ emission line possesses a
$\sim2000$ \kms-wide core and a redshifted tail that extends at least
another 800~\kms\ to the red.  However, the M\,87 absorption has
unknown \NHI, which makes it difficult to remove in order to determine
the intrinsic \Lya\ emission line properties.  Absorption has been
detected in two narrow components ($cz\approx980$ and 1330 \kms) in
\CaII\ H\&K and \NaI\ D by \citet{Carter92} and \citet{Carter97} and
in several other ions (\NaI\ D, \MgI, \MgII, \CaII\ H\&K, \MnII\ and
\FeII) in one broad component ($cz=1134\pm22$ \kms) by
\citet{Tsvetanov99} with FOS.  The variable absorption velocities
reported as well as the partial covering of the source suggested by
the \CaII\ and \NaI\ absorption, which have 1:1 doublet ratios but low
optical depths \citep{Carter97}, are strong indicators that variable
absorption exists in the nucleus of M\,87, perhaps due to cloud proper
motions.

On what is likely a slightly larger scale, HST/FOS off-nuclear optical
spectra and HST/Faint Object Camera (FOC) optical imaging discovered
an ionized gas disk with Keplerian motion \citep{Ford94,Ford99} that
connects to non-Keplerian blueshifted emission-line filaments
primarily to the NW of the nucleus.  \citet{Ford99} speculate that the
observed nuclear absorption lines found by \citet{Tsvetanov98} are
kinematically related to this filamentary gas since these are
blueshifted with respect to the nucleus as well. \citet{Macchetto97}
used the spectroscopic capabilities of the Faint Object Camera (FOC)
on HST to determine the rotation curve of the [\ion{O}{2}] emitting
gas in the nuclear disk which better determined the mass of the SMBH.
Relevant to this study, no [\ion{O}{2}] forbidden-line emission was
found within 0\farcs07 (5 pc) of the nucleus. Similar long-slit, optical
spectra obtained with the {\it Space Telescope Imaging Spectrograph}
(STIS) were obtained by \citet{Walsh13} to refine the SMBH mass
estimate.

Due to all these potential complications and uncertainties, we elected
to deconvolve the emission and absorption components in \Lya\ using an
{\it a priori} approach.  The \Lya\ profile is characterized by a
strong emission peak at $v_{\rm LSR}=1260$ \kms\ with a dereddened
peak amplitude of $8\times10^{-14}$ \flux.  This is $\sim6$ times
brighter than observed with FOS by \citealt{Sankrit99} on 1997, 23
January, and nearly $10\times$ brighter than on 1997, 18 January
(although these new observations are through a larger aperture;
2\farcs5 with COS compared to 0\farcs26 with FOS; see Section 3.1.2).
Weaker emission can be seen extending over the range
$1207<\lambda<1232$ \AA\ and thus blended with the Milky Way DLA. The
emission line profiles in the three COS spectra shown here, especially
the \Lya\ line in M\,87, are far too broad to be affected by any
spectral smearing caused by the marginal spatial resolution in the COS
aperture (see Section 3.4).

We model the \Lya\ emission profile as three emission components
(``blue'', ``main'' and ``red''), a damped Galactic \Lya\ absorber at
$v_{LSR}\sim0$, a weak, broad absorption component to accomodate
possible absorption in M\,87, and a linear far-UV continuum flux.  The
detailed flux model includes fifteen fitting parameters (three for
each Gaussian emission component, three for the M\,87 absorber, two
each for the Galactic DLA and the non-thermal continuum of M\,87),
each allowed to vary around a reasonable range.  Despite this
complexity, a remarkably robust fit is determined ($\bar\chi^2=0.92$)
even when fit parameters are allowed to vary over a broad range of
plausible values.  Best-fit quantities and $1\sigma$ fitting
uncertainties are given in Table~2 and shown overplotted on the data
in the left-hand panels of Figure~\ref{fig:m87lines}.

Blue and red wings are apparent in the emission profile and these are
best fit with two additional Gaussian components with centroids at
$1211.93\pm0.47$ \AA\ and $1228.00\pm0.07$ \AA, respectively, modified
by the Galactic DLA.  If \Lya\ emission, these correspond to $v_{\rm
LSR}=-922$ \kms\ and $v_{\rm LSR}=+3040$ \kms.  We note that the blue-
and red-wings of the profile {\bf cannot} be fitted with a single very
broad emission component at $v_{\rm LSR}\sim1200$ \kms; two Gaussian
profiles offset from the main, central component by $\sim1000-2000$
\kms\ are required.  Thus, if the blue and red emission are both
\Lya\, they are approximately symmetrically-placed around the main
\Lya\ line.  We interpret the red component as \Lya\ emission
redshifted by $\sim1800$ \kms\ with respect to the main \Lya\ emission
in agreement with \citet{Sankrit99}.  The maximum redshifted velocity
observed in this broad component is $\sim3000$ \kms\ relative to the
main \Lya\ line.

The blue side emission is possibly either \SiIII\ 1206 \AA\ emission
at $v_{\rm LSR}=+1540\pm140$ \kms\ or \Lya\ emission blueshifted at
$-922$ \kms\ LSR or $-2185$ \kms\ with respect to the main \Lya\
emission peak.  The larger uncertainty in the velocity centroid of
this line is due to the presence of the Galactic DLA, making this
component's position and profile very dependent upon the exact assumed
DLA column density (see below).

The \SiIII\ interpretation is consistent with the velocity and line
width of emission seen in \CII\ (see below), albeit with significant
uncertainty.  However, the velocity centroid of the proposed \SiIII\
emission is almost 300 \kms\ {\em greater} than the strong \Lya\
emission centroid, casting doubt on this identification.  Further, if
this emission is \SiIII, it would be much stronger than what is seen
in any other AGN.  In a typical Seyfert like Mrk\,817
\citep{Winter11}, a \SiIII\ line as luminous as the blue component in
M\,87 would be visible as an asymmetry in the line profile of \Lya.
This is not observed.  Further, the FUV spectum of the prototypical
narrow-line Seyfert\,1 galaxy, I\,Zw\,1 has only a reported marginal
detection of \SiIII\ emission at a level ten times weaker than the
feature we observe in M\,87. We conclude that this feature is unlikely
to be \SiIII.

If this emission is interpreted as \Lya, it has a velocity of $-922$
\kms\ LSR and a blueshifted velocity of $-2185$ \kms\ relative to the
\Lya\ emission peak, making it reasonably symmetrical in both velocity
offset and luminosity to the redshifted \Lya\ emission line which is
not affected by the Galactic DLA and has been reported previously.

A dip near the peak of the main component of the \Lya\ emission line
is well-fit as a very broad \Lya\ absorption line ($b\approx250$ \kms)
at $v_{\rm LSR}=1490$ \kms\ ($v=+230$ \kms\ relative to the main line
centroid).  While similar to the broad, low-ion detections seen by
\citet{Tsvetanov99}, this absorption component is at a very different
heliocentric velocity ($\approx+350$ \kms\ relative to the
\citet{Tsvetanov99} detections).  This absorption lies on the red side
of the \Lya\ peak compared to the blue side absorption seen in the
epoch 1997 FOS spectrum.

The Galactic DLA absorption we fit has a derived $\log N\rm_{HI}
(cm^{-2})=20.14$, quite close to but slightly less than the Galactic
$\log N\rm_{HI} (cm^{-2})=20.34-20.53$ value inferred by
\citet{Sankrit99}. This value is also close to but slightly larger
than the total $N_H$ found by \citet{DiMatteo03} from a continuum fit
to the X-ray emission at the M\,87 nucleus. If we constrain the
Galactic DLA \NHI\ to be between the two Sankrit values, our best-fit
solution is driven to the minimum value (i.e., $\log
N\rm_{HI}(cm^{-2}) =20.34$) and the intrinsic main \Lya\ emission
component would increase by a factor of 8\% in flux. However, the blue
emission is much more affected by the Galactic DLA increasing; its
line-flux increases by nearly a factor of two and its emission
centroid shifts to $1214.04\pm0.16$ \AA.  At nearly 1.5 \AA\ redward
of its location in the unconstrained model (Table~2,
Figure~\ref{fig:m87lines}), it is even more unlikely to be
\SiIII\ 1206\AA\ and much more likely to be a blue-shifted \Lya\
velocity component at $v_{lsr}=-400$ \kms\ or $v=-1670$ \kms\ with
respect to the main \Lya\ emission.

\begin{figure}
  \epsscale{1.1}\plotone{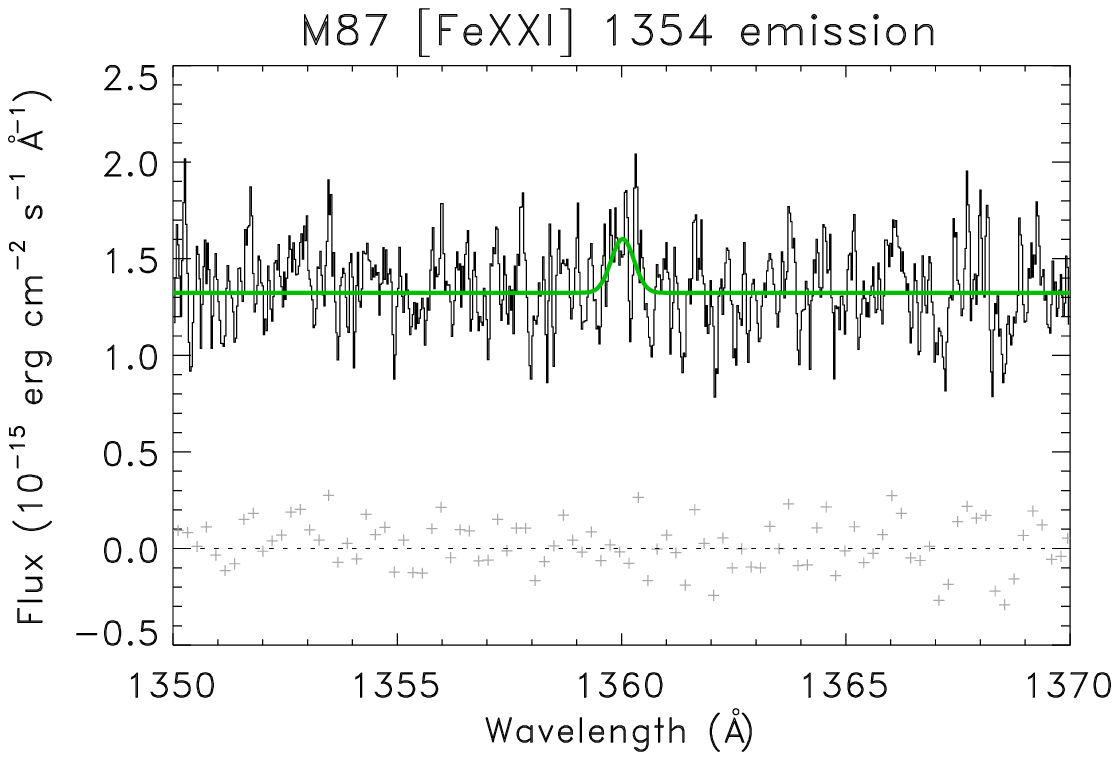}
  \caption{A $\sim5\sigma$ emission feature at 1360 \AA\ is consistent
  with redshifted [Fe\,XXI] $\lambda=1354.1$ emission.  A Gaussian fit
  to this feature gives $F_{\rm FeXXI}=(1.8\pm0.8)\times10^{-16}$ erg
  cm$^{-2}$ s$^{-1}$, $v_{lsr}=1316\pm29$ \kms, and $FWHM=130\pm70$
  \kms.}\label{fig:m87fe21}
\end{figure}

\subsubsection{Metal ion emission lines} 

In addition to the strong \Lya\ emission, there are a number of weak
emission features in the M\,87 spectrum.  As discussed above, the
emission feature near to a potential \SiIII\ in M\,87 was fitted
concurrently with \Lya\ but is interpreted as blushifted \Lya, not
\SiIII.  Peaks at $\sim1244$ \AA\ and $\sim1340$ \AA\ are consistent
with \NV\ doublet and \CII\ emission respectively.  Best fit
parameters for these lines are given in Table~2.

The \CII\ 1334.5 \AA\ emission line is modeled as a single Gaussian
along with a $v_{\rm LSR}\approx0$ Galactic \CII\ absorption line. The
Galactic \CII\ line is fitted with the parameters $v_{\rm LSR}=
+26\pm11$ \kms, $b=80\pm16$ \kms, $\log\,N (\rm cm^{-2})=15.0\pm0.1$.

The \NV\ doublet ($\lambda=1238.8$, $1242.8$) was fitted assuming an
identical velocity centroid and FWHM parameters for the two lines, but
letting the relative line strength vary from the nominal 2:1,
optically thin doublet ratio to an optically thick 1:1 ratio.

A low-contrast peak near 1400 \AA\ is consistent with a blended
complex of the \ion{Si}{4} $\lambda=1393.8$, $1402.7$ doublet and
\ion{O}{4}] $\lambda\approx1403$~\AA\ emission.  However, given the
low data quality near the edge of the COS/G130M detector and the
complicated nature of the required model, no fit was attempted.

\citet{AndersonSunyaev16} recently reported the tentative detection of
\FeXXI\ $\lambda=1354.08$ emission from M\,87.  A careful examination
of the data around $\lambda=1360$ \AA\ reanalyzed using the techniques
of \citet{Danforth16} shows a weak ($\sim5\sigma$) feature consistent
with [\ion{Fe}{21}] emission (Figure~\ref{fig:m87fe21}).  We measure
$F_{\rm [FeXXI]}=(1.8\pm0.8)\times10^{-16}$ erg cm$^{-2}$ s$^{-1}$,
$v_{lsr}=1316\pm29$~\kms, and $FWHM=130\pm70$ \kms.  The line flux
value quoted here is consistent with the upper limit of
($4\times10^{-16}$~erg~cm$^{-2}$~s$^{-1}$) quoted by
\citet{AndersonSunyaev16}.  While their estimate of line-width
$\approx290$~\kms\ is much higher than we obtain, it is based on a
marginal detection.  A value of $FWHM=130\pm70$~\kms\ is both larger
than a thermal width of $\sim50$ \kms\ for this species and provides a
first estimate of the turbulence of the hot gas close to the SMBH of
$\approx100$~\kms \citep{AndersonSunyaev16}.  New exposures of
comparable integration time to the values in Table~1 will make a
definitive measurement of turbulence in the accretion flow. This will
be reported at a later time.

\subsubsection{Line and continuum variability}

\citet{Sankrit99} reported on two epochs of nuclear FOS observations in
January 1997 separated by five days in which the UV continuum remained
approximately constant while the \Lya\ emission line flux increased by
a factor of $\approx1.6$. However, these authors caution that this
apparent variability may be due to a slight displacement of the FOS
aperture. Earlier FOS spectra from late 1993 into 1995 reported on by
\citet{Tsvetanov98,Tsvetanov99} found that the NUV $+$ optical
continuum (2000-4000\AA) was variable by a factor of $\sim2$ in 2.5
months, showed smaller variability on a timescale of weeks but no
change over a one day period.  Multi-epoch NUV imaging compiled by
Madrid (2009; 30 epochs) and Perlman \etal\ (2011; 20 epochs) in more
recent years finds $\sim20$\% continuum variability over timescales of
months to years.  Variability at the few percent levels are possible
on shorter time periods of 2-4 weeks.

While we can not be definitive about the \Lya\ emission line
variability in our observations relative to the FOS observations due
to the much larger COS aperture, it is likely that at least some of
the factor of 6--10 flux difference is due to intrinsic variability.
In subsection 3.4 below we use the limited spatial resolution of the
COS spectroscopy to determine that both the nuclear \Lya\ line and the
UV continuum are spatially extended by $\sim1$\arcsec. In the STIS
spatially-resolved spectroscopy of \citet{Macchetto97} and
\citet{Walsh13} the observed flux of the optical emission lines is
confined primarily to the inner 0\farcs5. If the spatial extent of
\Lya\ is similar, then most of the luminosity of \Lya\ is coming from
a region not too much larger than the FOS aperture. While proof of
\Lya\ variability awaits future COS spectroscopy for accurate flux
comparisons through the same aperture, an order of magnitude increase
seems quite extreme, suggesting that at least some of the FOS-COS flux
difference is intrinsic variability.  If the line variability over
long time intervals (months to years) is modest ($\sim20$\%), similar
to the observed continuum variability, the \Lya\ line flux could be
responding to earlier continuum variations (a.k.a. ``reverberation'').

However, the red wing seen in the \citet{Sankrit99} FOS spectrum and
in Figure~1 is the most variable portion of the emission profile and
is unaffected both by the Galactic DLA and the reported locations of
the absorption in M\,87.  Therefore, this emission component must
possess significant intrinsic variability.  The potential variability
of the blue wing emission is not yet known until we obtain new epochs
of COS FUV spectroscopy. The only constant emission or absorption
component in the vicinity of the M\,87 \Lya\ emission line is the
foreground Galactic absorption.

If the \Lya\ line variability observed in two FOS epochs is intrinsic,
a very small upper limit on the size of the \Lya\ emitting region can
be set by $c\tau\sim10^{16}$~cm \citep[$\sim10$ Schwarzschild radii
for a $4\times10^9~M_\sun$ SMBH; ][]{Sankrit99}.  But this
interpretation is inconsistent both with the rather narrow line width
of \Lya\ and with the observed spatial distribution of the emission
(see Section 3.4). This is simply closer to the SMBH than can be
accommodated by the emission line widths. A model in which the \Lya\
line flux changes due to a change in the amount of cloud mass in the
\Lya\ emitting region (i.e., variable accretion in a density-bounded
region, as was suggested in Paper~1) is also ruled out since any
variability would occur on a timescale over which clouds move in and
out of the emitting region at $v\leq1000$ \kms.  This model would
suggest variations only on a several year or longer timeframe, not on
the short timescales observed. Apparent emission line variability due
to variable absorption is a possibility as long as the line
variability is modest (few percent level). Again, future FUV
spectroscopy with COS can settle the important question of \Lya\
variability through comparison to the spectrum presented here.

\begin{figure} 
   \epsscale{1}\plotone{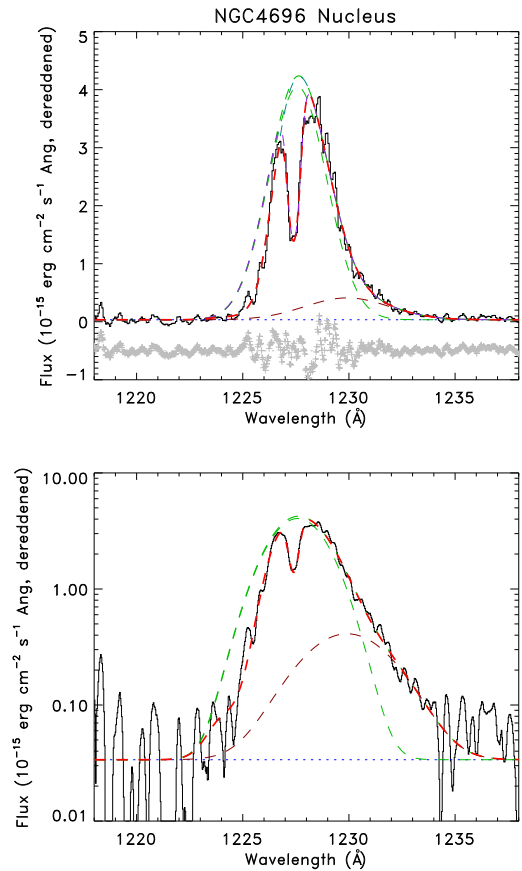}
   \caption{NGC\,4696 shows asymmetric \Lya\ emission with what
   appears to be a narrow absorption feature superimposed near the
   peak.}\label{fig:ngc4696}
\end{figure} 

\begin{figure}
  \epsscale{1}\plotone{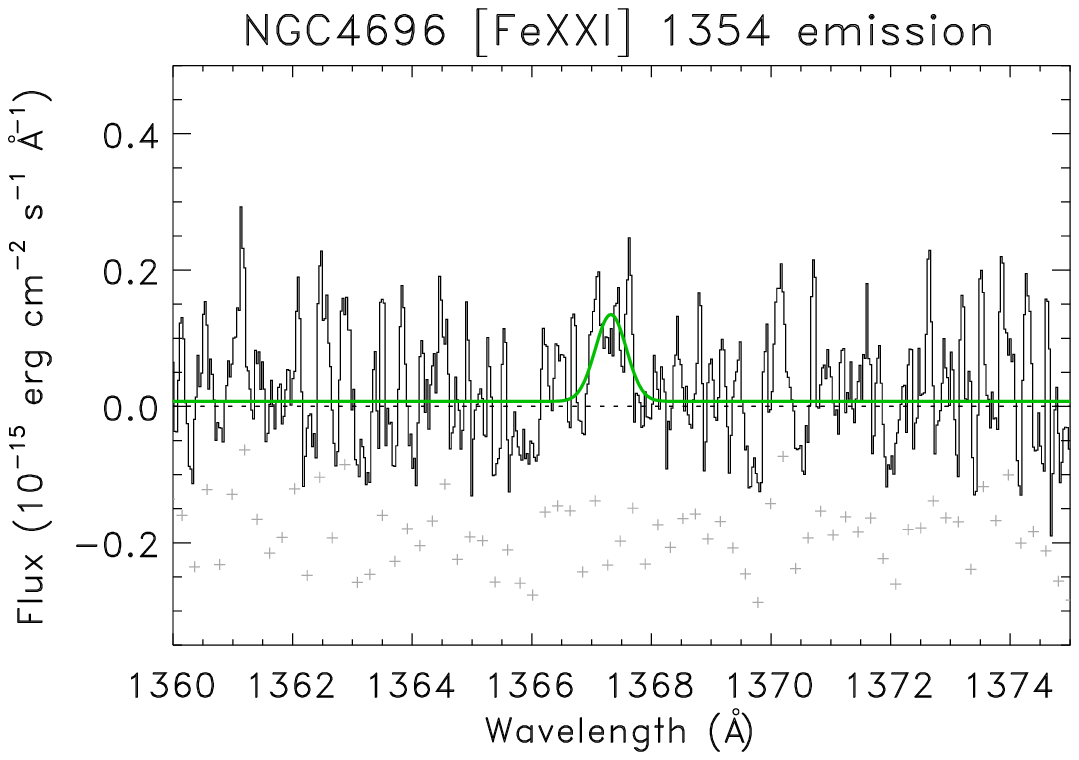}
  \caption{A $\sim2\sigma$ emission feature at 1367 \AA\ is consistent
  with redshifted [Fe\,XXI] $\lambda=1354.1$ emission.  A Gaussian fit
  to this feature gives $F_{\rm FeXXI}=(8\pm5)\times10^{-17}$ erg
  cm$^{-2}$ s$^{-1}$, $v_{lsr}=2931\pm38$ \kms, and $FWHM=135\pm83$
  \kms.}\label{fig:ngc4696_fe21}
\end{figure}

\begin{deluxetable*}{lccl}   
\tabletypesize{\footnotesize}   
\tablecolumns{4}    
\tablewidth{0pt}    
\tablecaption{NGC\,4696 Line Profile Fit Parameters}   
\tablehead{ 
 	\colhead{Quantity}   &
        \colhead{Comp. 1}&  
        \colhead{Comp. 2}&  
	\colhead{Unit}    }  
\startdata  
\cutinhead{\Lya\ emission components} 
 Centroid 	   &$ 1227.59\pm0.30 $&$ 1230\pm7$ & \AA \\ 
          	   &$ 2940\pm70$      &$\sim3500$  & \kms \\ 
 FWHM     	   &$3.2\pm0.8$       &$\sim5$     & \AA \\ 
  	  	   &$790\pm200$       &$\sim1200$  & \kms \\ 
 $I_{\rm Ly\alpha}$&$14\pm9$          &$\sim2$     & $10^{-15}\rm~erg~cm^{-2}~s^{-1}$ \\ 
 $L_{\rm Ly\alpha}$&$2.3\pm1.6$       &$\sim0.3$   & $10^{39}\rm~erg~s^{-1}$\tnma \\ 
 continuum	   &$3\pm1$           &            & $10^{-17}\rm~erg~cm^{-2}~s^{-1}~\AA^{-1}$ \\ 
\cutinhead{\Lya\ absorption components} 
 Centroid 	   &$2893\pm10$       &$\sim2280 $ & \kms \\ 
 $\log\,N_{\rm HI}$& $14.10\pm0.08$   &$14.7\pm0.6$& \\ 
 $b$      	   &$87\pm11 $        &$270\pm140$ & \kms  
   	\enddata 
  \tablenotetext{a}{assuming $D=37.6$ Mpc} 
\label{tab:ngc4696} 
\end{deluxetable*} 

\begin{deluxetable*}{lccl}  
\tabletypesize{\footnotesize}  
\tablecolumns{4}   
\tablewidth{0pt}   
\tablecaption{Hydra\,A \Lya\ Emission Line fits}  
\tablehead{
	\colhead{Quantity} & \colhead{Comp. 1}& \colhead{Comp. 2}&  \colhead{Unit} }
\startdata 
	centroid   &$ 1281.85\pm0.02$&$1281.80\pm0.02$& \AA \\
 	           &$-130\pm10      $&$ -140\pm10    $& \kms~\tnma \\
 	FWHM       &$1.44\pm0.09    $&$ 2.61\pm0.08  $& \AA \\
 	           &$ 336\pm21      $&$  610\pm19    $& \kms \\
 $I_{\rm Ly\alpha}$&$16.3\pm2.9     $&$ 29.5\pm2.9   $& $10^{-15}\rm~erg~cm^{-2}~s^{-1}$ \\
 $L_{\rm Ly\alpha}$&$ 112\pm20      $&$  203\pm20    $& $10^{39}\rm~erg~s^{-1}$~\tnmb \\
 	continuum  &$0.58\pm0.02    $&                & $10^{-15}\rm~erg~cm^{-2}~s^{-1}~\AA^{-1}$ \\
 	slope      &$  11\pm2       $&                & $10^{-18}\rm~erg~cm^{-2}~s^{-1}~\AA^{-2}$  
 \enddata 
  \tablenotetext{a}{with respect to a systemic redshift of $z=0.054878$}
  \tablenotetext{b}{assuming $D=240$ Mpc}  
  \label{tab:hydraa}
\end{deluxetable*}

\medskip
\subsection{NGC\,4696} 

NGC\,4696 shows an asymmetric \Lya\ emission profile at 1227.6 \AA\
($z=0.0090$, $cz=2710$ \kms) with what appears to be a narrow
absorption line superimposed at 1227.3 \AA.  There is a very low level
of continuum flux ($F<10^{-16}$ \flux); the continuum level listed in
Table~3 was obtained by averaging the flux in continuum pixels between
$1218<\lambda<1240$ \AA.  The component structure of this emission
line is unclear.  We interpret the narrow absorption as \Lya\ at the
same redshift as NGC\,4696. There are no Galactic, $z=0$ absorption
lines expected at this wavelength and the pathlength to NGC\,4696 is
short enough that no intervening, intergalactic absorption lines are
expected. The simplest model--a single Gaussian emission component and
a single Voigt absorption profile--does not give satisfactory results.
Adding a second, weaker emission component helps account for the red
wing of the emission, but the peak is still poorly fit.  The steep
blue edge of the emission suggests the presence of additional
absorbing gas slightly blueshifted from the systemic \Lya\ of
NGC\,4696, possibly outflowing from the nucleus or associated with the
host galaxy itself.  If a single, broad absorption line is included in
the fit, the profile model fits the data well.  However, we caution
against overinterpretation of this fit, especially the inferred
blue-side absorber.  The emission may be intrinsically non-Gaussian or
the absorption profile may be more complicated than a single, simple
Gaussian component.  Fit parameters are given in
Table~\ref{tab:ngc4696}.

\citet{AndersonSunyaev16} found a very low-significance emission
feature they interpreted as \FeXXI\ associated with NGC\,4696.  We
confirm this possible detection in our more sophisticated reduction of the data
(Figure~\ref{fig:ngc4696_fe21}) at a significance level of
$\sim2\sigma$.  The possible emission line is fitted by a Gaussian
with $FWHM=135\pm83$ \kms\ and $v_{lsr}=2931\pm38$ \kms.  Our fitted
flux of $F_{\rm FeXXI}=(8\pm5)\times10^{-17}$ erg cm$^{-2}$ s$^{-1}$
is consistent with the Anderson \& Sunyaev 90\% upper limit of $F_{\rm
FeXXI}\le2.2\times10^{-16}$ erg cm$^{-2}$ s$^{-1}$.

\subsection{Hydra\,A}

\begin{figure}
   \epsscale{1}\plotone{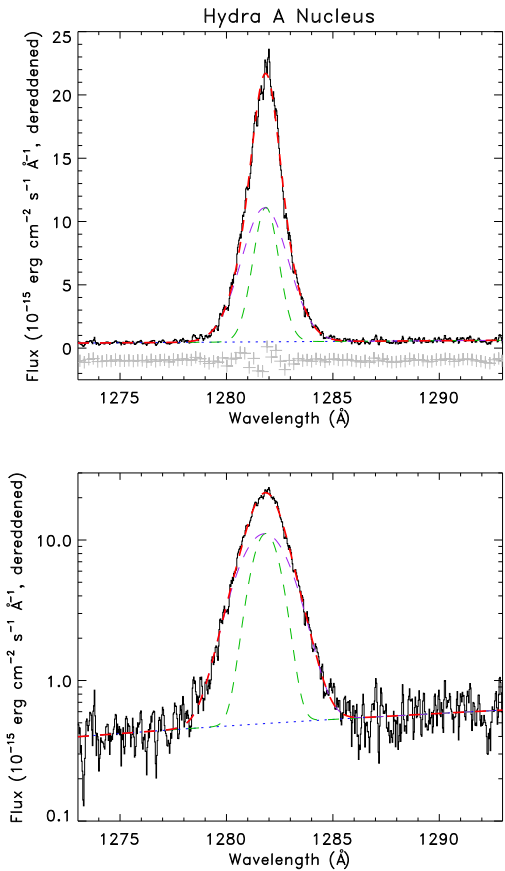}
   \caption{Emission line profiles and model fits for Hydra\,A.  \Lya\
   emission can be well fitted with a pair of Gaussian components.}
   \label{fig:hydraa}
\end{figure}

The far-UV spectrum of Hydra\,A shows a relatively narrow \Lya\
emission peak at close to the systemic velocity.  It is well-fit with
a broad-plus-narrow pair of Gaussian emission components and a
first-order linear continuum (Figure~\ref{fig:hydraa}).  Some
structure in the residual suggests that there may be additional
components, but nothing obvious is seen.

Recent HST images of Hydra\,A in the ultraviolet by \citet{Tremblay15}
show that the nucleus is obscured by a dust disk viewed edge-on.  Thus
one would expect the nuclear FUV continuum and line emission to be
suppressed or absent entirely.  It is therefore surprising that a FUV
continuum is present, the \Lya\ line emission is many times more
luminous than either M\,87 or NGC\,4696 (Tables~2, 3), and that no
narrow absorption line is seen superimposed on the \Lya\ emission as
is seen in the other two cases (Figures~1, 3).  We conclude that the
nuclear region must therefore not be obscured by the dust disk in
Hydra\,A.

In addition to the usual Galactic, $z=0$ absorption lines, the
Hydra\,A spectrum shows hints of \CII\ 1334.5, \OI$+$\SiII\ 1304, and
\SiIII\ 1206.5 absorption at the redshift of Hydra\,A.  It is possible
that this absorption arises in the dust disk bisecting the galaxy
\citep{Tremblay15}.  However, no \HI\ absorption is seen at the
systemic redshift, so modeling of these absorption lines would require
data of much higher quality.

No metal-ion emission is seen including \FeXXI.

\subsection{Limited angular resolution information from HST/COS}

While $HST$/COS was not designed to provide high angular resolution
imaging spectroscopy, it has been demonstrated that limited spatial
information can be obtained from an analysis of a target's
cross-dispersion profile \citep{France11}.  The angular resolution of
COS in the G130M mode is $0.8-1.1$\arcsec.  In order to evaluate the
spatial extent of the \Lya\ emission line region at $\approx1\arcsec$
resolution, we compared the angular profiles of the three FR\,1
galaxies with 1) a known extended emission source and 2) a known point
source.  HST orbits in a diffuse cloud of neutral and ionized atoms;
the ``air glow'' from the atomic recombination and resonant scattering
of solar photons in this cloud produces a uniform background of
hydrogen and oxygen emission that fills the COS aperture and
approximates a filled-slit observation of an astronomical target.  For
point source comparison, we downloaded spectra of a well-studied COS
calibration target, WD\,0308$-$565.  We analyzed observations from the
longest central wavelength settings available from each target ({\sc
cenwave}$=1318$ and $1327$) as these grating settings have the
smallest intrinsic cross-dispersion heights
\citep{Roman_duval13_COS_ISR}.

\begin{figure}
 \epsscale{1}\plotone{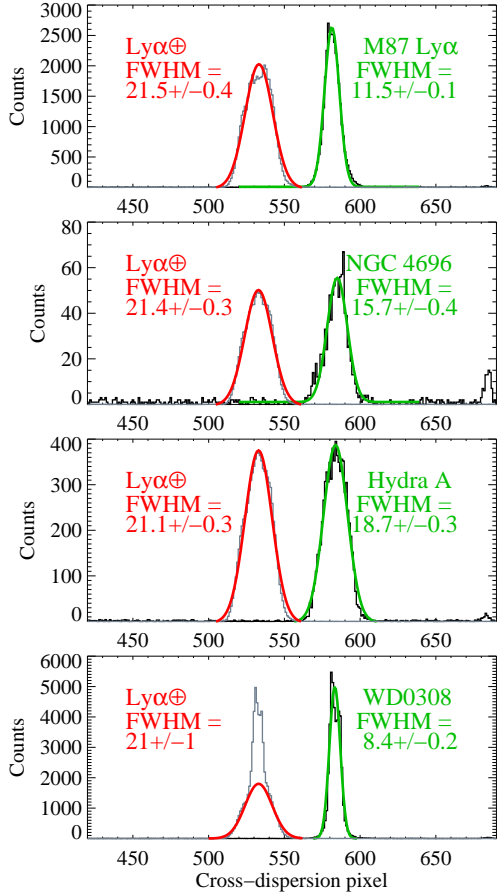}
 \medskip\medskip \caption{The profile in the cross-dispersion
 direction for our three FR\,1 nuclei and calibration source WD\,0308.
 Each panel shows both the object spectrum and the geocoronal \Lya\
 airglow profile (which fills the aperture).  The FWHM values of the
 three FR\,1s are intermediate between the WD (a point source) and the
 \Lya\ airglow (which fills the aperture in all cases) implying that
 their FUV emission is partially resolved by COS.}  \label{fig:xdisp}
\end{figure}

Figure~\ref{fig:xdisp} shows the cross-dispersion profiles of the
\Lya\ emission lines of the three FR\,1 galaxies (top three panels),
the FUV continuum of WD\,0308 (bottom panel), and the \Lya\ airglow
spectra measured simultaneously.  The galactic/stellar profiles were
extracted over a 400 dispersion-direction pixel region ($\approx4$
\AA) centered on the source \Lya\ emission line (offset into the
continuum at 1247.8 \AA\ for WD\,0308).  The geocoronal \Lya\ airglow
profile was fitted with a Gaussian (red) and shown offset from the
central cross-dispersion profile of the targets by $-50$ pixels.
Gaussian fits to the galaxy/star cross-dispersion profiles are shown
in green.  The Gaussian FWHMs (in pixels, recall 1 pixel
$\approx0.11$\arcsec\ for COS/G130M) are shown in the legends.  The
y-axis label shows the total counts per cross-dispersion extraction
region in each of the targets, with the geocoronal profile scaled to
the peak of the target flux.  For the FR\,1 galaxies, the geocoronal
\Lya\ line-height is always between $21-22$ pixels FWHM
($2.3-2.4$\arcsec), consistent with the expected instrumental profile
for a filled aperture and the non-uniform primary science aperture
response function of COS.  The WD\,0308 continuum is bright enough
that it has comparable flux to the airglow line, thus its geocoronal
line width must be fitted with a pair of Gaussians.  The FWHM of the
white dwarf cross-dispersion profile is $8.4\pm0.2$ pixels
(0.92\arcsec), consistent with the expected cross-dispersion profile
for a point source.  The broader component has FWHM$=21\pm1$ pixels
(2.3\arcsec), consistent with a filled-aperture (bottom panel,
Figure~\ref{fig:xdisp}).

Contrasting the filled-aperture geocoronal observations and the point
source white dwarf observations with the cross-dispersion heights of
the FR\,1 galaxies, we see that all three galaxies are in an
intermediate category: they are clearly extended emission sources, but
do not fully fill the COS aperture.  For M\,87, we find a
cross-dispersion line height of $11.5\pm0.1$ pixels (1.27\arcsec),
with $18.7\pm0.3$ pixels (2.06\arcsec) and $15.7\pm0.5$ pixels
(1.73\arcsec) for Hydra\,A and NGC\,4696, respectively.  We conclude
that the \Lya-emitting regions for these FR\,1 galaxies are indeed
extended, though do not fill the aperture.

A similar analysis of a region of the FUV continuum away from the
galactic \Lya\ emission signal shows an interesting result: the FUV
continua in M\,87 and Hydra\,A are also extended (that is, not
consistent with a point-like AGN alone) and the FUV continuum in M\,87
shows an angular extent $20-40$\% larger than the \Lya\ emission
region (FWHM$=14-16$ pixels over a range of continuum wavelengths and
extraction region sizes).  NGC\,4696 does not have sufficient FUV
continuum flux for a cross-dispersion profile to be reliably measured.

\begin{figure}
  \epsscale{1}\plotone{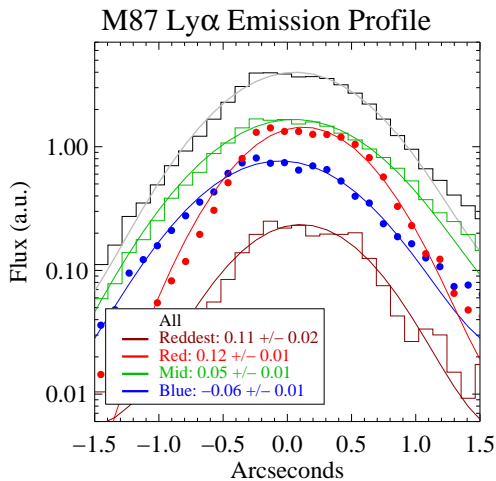}
  \caption{The spatial (cross-dispersion) profile of the M\,87 \Lya\
  emission line is velocity-dependent consistent with material
  entrained in the outflowing jet and counter-jet.}  \label{fig:tilt}
\end{figure}

Furthermore, in M\,87 the \Lya\ core emission, blue wing emission, and
red wing emission are centered at slightly different locations in the
aperture (see Figure~\ref{fig:tilt}).  While slight, these offsets are
consistent with \Lya-emitting material outflowing along the direction
of the jet (blue wing is in the direction of the jet; red wing in the
direction of the counterjet).

The inferences on the spatial extent and location of the line and
continuum emission in M\,87 can be tested by obtaining long-slit FUV
spectra of the nucleus at position angles along the jet and
perpendicular to the jet using HST/STIS.  The results of these
observations (approved for HST cycle 24) will be reported in a future
publication.

\subsection{BL\,Lac Objects} 

\begin{figure} 
  \epsscale{.9}\plotone{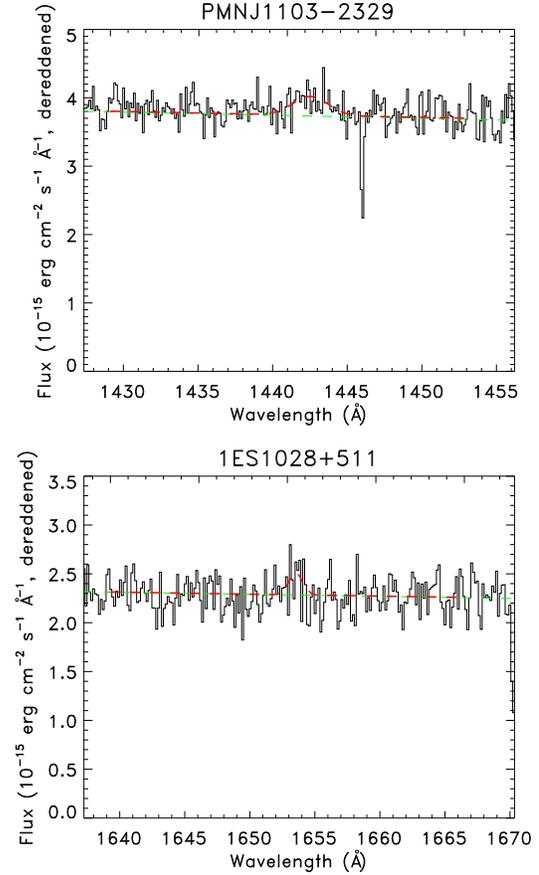}
  \caption{Weak \Lya\ emission for two unpublished BL\,Lac objects.}
\label{fig:new_bllacs} 
\end{figure} 

Several BL\,Lac objects in the literature show weak \Lya\ emission:
Mrk\,421, Mrk\,501, and PKS\,2005$-$489 (Paper~1) and more recently
H\,2356$-$309 \citep{Fang14}.  To these we add new detections of weak
\Lya\ emission in PMN\,J1103$-$2329 and 1ES\,1028$+$511.  As with
Paper~1, we fit power-law continua to the spectra and model the
\Lya\ emission profiles as single, Gaussian components.

Neither PMN\,J1103$-$2329 nor 1ES\,1028$+$511 show high-significance
\Lya\ emission features like the FR\,1s, nor even as large an
equivalent width as the previously-published BL\,Lac objects with
\Lya\ emission (Paper~1).  Nevertheless, a Gaussian fit in the
expected location of the systemic \Lya\ shows weak, broad, emission
with significance levels of $\sim6\sigma$ and $\sim3\sigma$ for
PMN\,J1103$-$2329 and 1ES\,1028$+$511, respectively
(Figure~\ref{fig:new_bllacs}).  

Of the seven BL\,Lac objects with well-known redshifts and high S/N
FUV spectra, six show intrinsic \Lya\ emission.  Weak, relatively narrow \Lya\ emission
appears to be a generic feature of both BL\,Lac objects and their
parent population FR\,1 radio galaxies.

\medskip
\section{Discussion} 


The similar \Lya\ luminosities of FR\,1s and BL\,Lac objects is
additional evidence that these two classes are the same type of object
seen from different viewing angles. Specifically, the BL\,Lacs
(Mrk\,421, Mrk\,501 and PKS\,2005$-$489) and the FR\,1 (Hydra\,A) are
at comparable redshifts and have comparable \Lya\ luminosities (see
Table~\ref{tab:lyaproperties}).  But the line luminosities for the
observed sources vary by two orders of magnitude, an unexpected result
if the ionizing continuum and distribution of \Lya\ emitting clouds in
these objects are both isotropic.  The pure FR\,1s in
Table~\ref{tab:lyaproperties}, M\,87 and NGC\,4696, have at least an
order of magnitude less \Lya\ luminosity than any of the other AGN
which have at least some beamed radiation coming in our direction.
Therefore, the beamed radiation must be considered in modeling the
\Lya\ emission, which makes the emission mechanism and its location(s)
a great deal harder to model in these objects.
  
\begin{deluxetable*}{llccccc}
  \tabletypesize{\footnotesize}  
	\tablecolumns{7}   
	\tablewidth{0pt}   
	\tablecaption{\Lya\ Emission properties of FR\,1 and BL\,Lac AGN}  
	\tablehead{
  	  \colhead{Target}   & 
	  \colhead{$z_{\rm AGN}$} &
	  \colhead{SL} & 
	  \colhead{FWHM \tnma}& 
	  \colhead{$L_{\rm Ly\alpha}$ \tnmb} & 
	  \colhead{Instrument} & 
	  \colhead{Notes}    \\
	  \colhead{}   & 	
	  \colhead{} & 
       	  \colhead{($\sigma$)} & 
          \colhead{(\kms)}&
          \colhead{$\rm 10^{40}~erg~s^{-1}$} & 
	  \colhead{} & 
	  \colhead{}    }
   \startdata      
 \cutinhead{FR\,1 AGN}
 	M\,87             & 0.0022 &    &$1240\pm20$ &$0.52\pm0.04 $&HST/COS & this work \\
  	NGC\,4696         & 0.00987&    &$790\pm200$ &$0.26\pm0.16 $&HST/COS & this work \\
  	Hydra\,A          & 0.05488&    &$610\pm20 $ &$32\pm3 $     &HST/COS & this work \\
       \cutinhead{BL\,Lac Objects}				        
       	PMN\,J1103$-$2329 & 0.1847 & 6.2&$ 580:$     &$4.9\pm0.4 $  &HST/COS & this work \\
  	1ES\,1028$+$511   & 0.3607 & 3.3&$ 220:$     &$5.8\pm1.7 $  &HST/COS & this work \\
  	H\,2356$-$309     & 0.165  & \nd&$1340\pm320$ &$9.53\pm2.02$&HST/COS & \citet{Fang14} \\
  	Mrk\,421          & 0.0300 &  9 &$300\pm30$ &$2.37\pm0.22$  &HST/COS & Paper 1 \\
  	PKS\,2005$-$489   & 0.0710 & 15 &$1050\pm60$&$24.9\pm1.1$   &HST/COS & Paper 1 \\
  	Mrk\,501          & 0.0337 & 23 &$820\pm80$ &$5.2\pm0.3 $   &HST/FOS & Paper 1 \\
  	PKS\,2155$-$304   & 0.116  &$<4$&\nodata    &$       <11$   &HST/STIS& Paper 1 
      \enddata   
 \tablenotetext{a}{FWHM of strongest emission component only}   
 \tablenotetext{b}{total \Lya\ luminosity of all components} 
	\label{tab:lyaproperties}
 \end{deluxetable*}

In a Seyfert galaxy, the line emission comes from the integrated
emission of the broad and narrow emission line clouds while the continuum arises in an
accretion disk.  The low-level, non-thermal continuum emission in
FR\,1s suggests that an accretion disk is not present or at least does
not possess anywhere near the accretion rate seen in Seyferts
\citep{Prieto16}.  What, then, is the ionizing radiation source we
detect in recombining \HI?  And what is the distribution of the
emission line clouds?  Our COS spectra are suggestive of answers to
these questions but are not conclusive.

\begin{deluxetable*}{lcccl}
\tabletypesize{\footnotesize}
	\tablecolumns{5}   
	\tablewidth{0pt}   
	\tablecaption{\Lya\ overprediction factors}  
	\tablehead{
	  \colhead{Quantity} &
	  \colhead{M\,87} &
	  \colhead{Hydra\,A} &
	  \colhead{NGC\,4696} &
	  \colhead{Unit}}
\startdata
  $I_{1215}$\tnma     &$1.37\times10^{-15}$&$5.9\times10^{-16}$&$  4\times10^{-17}$&$\rm erg~cm^{-2} s^{-1} \AA^{-1}$\\
  $\alpha_\lambda    $&$   0.92           $&$    0.85         $&$    0.97         $&\\
  $I_{912}           $&$1.79\times10^{-15}$&$7.6\times10^{-16}$&$  5\times10^{-17}$&$\rm erg~cm^{-2} s^{-1} \AA^{-1}$\\
  ionizing flux       &$   0.0687         $&$    0.0273       $&$    0.0021       $&photons $\rm cm^{-2} s^{-1}$\\
  predicted $I_{Ly\alpha}$&$120 $          &$45$               &$  3.0$            &$\rm 10^{-14}~erg~cm^{-2} s^{-1}$\\
  observed $I_{Ly\alpha} $&$ 60 $          &$4.5$              &$  1.4$            &$\rm 10^{-14}~erg~cm^{-2} s^{-1}$\\
  OPF                 &$  2.0             $&$   10            $&$    2.1          $&  predicted/observed  
\enddata   
\tablenotetext{a}{Continuum flux at 1215 \AA.}
	\label{tab:opf}
\end{deluxetable*}

In Table~\ref{tab:opf} we use COS observations of the three FR\,1s to
determine the factor by which the ionizing continuum extrapolated from
our COS FUV spectra overpredicts the strength of \Lya\ emission: the
so-called over-prediction factor (OPF) of Paper~1.  The OPF is the
ratio between the Ly $\alpha$ luminosity predicted from the power-law
fit to the ionizing continuum, to that observed in COS spectra, and it
should be related to the Doppler factor $\delta = [\Gamma(1 - \beta
\cos \theta)]^{-1}$.  Whereas the BL\,Lac objects exhibit OPFs of
hundreds to tens of thousands, two of the three FR\,1s have OPFs only
slightly greater than unity.  In previous FOS observations of M\,87
\citep{Sankrit99} the \Lya\ luminosity was lower but the continuum
flux was at the same level as shown here, leading to OPF values
slightly less than unity inside the FOS aperture.  The OPF of Hydra\,A
is intermediate between the very nearby FR\,1s and the BL\,Lac objects
as might be expected if we are seeing a portion of the the beamed
continuum in that AGN.  A small OPF ($\sim3$) is also seen in the
prototypical Seyfert~1 galaxy Mrk\,817 so that the OPF values slightly
greater than one in Table~\ref{tab:opf} likely are consistent with
photo-ionization with modest covering factor for the emission-line
clouds (see Paper~1).

If the ionizing continuum producing the \Lya\ in these FR\,1s is the
result of beaming, then the ionizing continuum seen by potential \Lya\
emitting clouds should vary smoothly with angle away from the beaming
axis.  If \Lya\ emitting clouds exist isotropically around the
continuum source, then some clouds are illuminated by a much larger
ionizing continuum than we observe.  For example, it has been
estimated from population statistics \citep{Perlman93,UrryPadovani95}
that the half power-angle of the X-ray emission in BL\,Lacs is
$\sim25$\arcdeg, which produces a significant fraction of beamed continuum
at all angles relative to the beaming axis \citep[see plots in the
Appendix of][]{UrryPadovani95}. In this case the OPFs in
Table~\ref{tab:opf} could be significant underestimates; i.e., there
is much more ionizing continuum illuminating the near nuclear region
than the amount of \HI\ that can absorb it.  Again, anisotropic
distributions of ionizing flux and \Lya\ emitting clouds are suggested.

\subsection{M\,87}

While all three FR\,1 objects studied here show intriguing complexity,
more constraints on source size and geometry are available for M\,87
due to its proximity to us and its previous observations in the FUV by
HST.  In particular, M87 is the only one of these objects where the
jet viewing angle ($\approx 15^\circ$) is well constrained
\citep{Perlman11,Meyer13}.  Variability information is particularly
important and puzzling.  The \Lya-emitting gas cannot possibly be as
close in as $10^{16}$~cm (0.003 pc) to the SMBH as claimed by
\citet{Sankrit99} from their variability constraints.  The Schwarzschild
radius for the SMBH in M\,87 is $\sim10^{15}$~cm.  Even at
$10^{17}$~cm ($100~R_S$) the Keplerian speed is $>10^4$ \kms, so the
lines would be an order of magnitude broader than we observe if the
\Lya-emitting gas were even this close to the SMBH.  A more likely
location for the \Lya\ clouds is $\sim10^4~R_S$ ($\sim10^{19}$~cm;
$\sim3$~pc), the inner reaches of the narrow-line region and of the
observed gaseous disk. So, there is little chance that the
FOS-observed variability could be due to a very small source size. The
larger distance of $\sim3$~pc from the SMBH is consistent with the
observed angular extent of the \Lya\ components in the COS aperture as
well as the observed line-width, but leaves the large variability
amplitude over short time periods as a puzzle. It seems most likely
that the reported 5 day \Lya\ variability is spurious, due to a
mis-placement of the FOS aperture as suggested by \citet{Sankrit99}.
Currently there is no unambiguous indication for emission-line gas
closer than 3~pc, at much larger radii than where the BLR of Seyferts
exists.

The suppression of BLR emission in BL\,Lacs was first explored by
\citet{Guilbert83} who pointed out that the steep spectrum ionizing
continua seen in BL\,Lacs will inhibit the creation of cool, $10^4$~K
clouds that are typical of Seyferts and QSOs.  For this mechanism to
be operable in FR\,1s, this steep ionizing continuum should be
illuminating much of the potential BLR of these AGN.  Indeed, the
steep X-ray spectra of BL\,Lacs must be much less beamed (half
illumination angle $\sim1$ radian) than the radio and optical continua
based on their observed properties and source counts
\citep{UrryPadovani95} when selection is made in the X-ray versus
radio or optical bands.  Therefore, the presence of broad-line
emitting gas is {\bf not} expected in this class of AGN, as we
observe.

Unlike the narrow core emission in M\,87, the red and blue wings to
the \Lya\ emission are much broader and appear to be offset spatially
from the main \Lya\ emission along the jet axis (see Sec~3.4).  The
red wing emission also varies over the same observing epochs and is
very likely intrinsic variability since there is no observed
absorption in M\,87 at these wavelengths.  Given their
oppositely-trending spatial extents, the red and blue emission wings
are likely to arise in gas illuminated or shocked by the jet (blue
wing) and counter-jet (red wing).  Spatially-resolved STIS spectra can
confirm this suggestion.

\subsection{FR\,1 and BL\,Lac Accretion Powers}

In the three FR\,1s observed by HST/COS, the \Lya\ luminosities are
comparable to or somewhat less than what is expected in ``Case B''
recombination theory assuming unity covering factor.  In M\,87
previous epochs saw lower \Lya\ luminosities while the continuum
luminosities remained the same.  Therefore, at the highest \Lya\
luminosities observed, an ionization-bounded scenario is possible in
these AGN but, even then, the amount of gas must be close to the
amount needed to absorb all the available ionizing photons given the
OPF $\approx1$ values found here. If the line-emitting clouds are in a
density-bounded regime then, as illustrated in Paper~1 using the \Lya\
line luminosities seen in BL\,Lac objects, there is very little warm
gas in this region, $10^{-4}-10^{-5}~M_\sun$.  This amount is inferred
by assuming that there is one hydrogen atom for every \Lya\ photon
emitted.  A reasonable energy conversion rate of this mass (e.g.,
$\sim$1\%) finds a very low estimated accretion rate, which is
consistent with earlier estimates using a variety of methods
\citep{Kuo14, Russell15, DiMatteo16}.

It is also possible that the observed \Lya\ gas is {\em outflowing} rather
than infalling, which decreases the estimated accretion rate even further.  
This possibility is consistent with the limited red/blue wing
spatial information from the COS aperture (Section 3.4). It is
important to verify this assertion using the full HST spatial
resolution with STIS.

There could be a significant amount of much hotter, ``coronal'' gas in
the line-emitting region surrounding the cooler, \Lya-emitting clouds.  Using
the results from \citet{AndersonSunyaev16} on the detection of \FeXXI,
which we confirm here, an estimate of particle density of
$0.2\rm~cm^{-3}$ can be derived for a temperature of $T\approx10^7$~K
based on the presence of the \FeXXI\ feature. Assuming this hot
coronal gas fills the region around the SMBH out to $\sim3$~pc, this
phase contains $\sim0.3~M_\sun$ of gas. These values are consistent
with the recent re-analysis of {\it Chandra} X-ray images of M\,87
conducted by \citet{Russell15}.  Although we have no good, direct
measurement of the density of the \Lya-emitting clouds, if
the coronal gas is in pressure equilibrium with these clouds, their
densities are a few $\times10^{-4}\rm~cm^{-3}$.  It is reasonable to
suspect that the \Lya-emitting cloud ensemble is gas which has
condensed out of this hotter phase in either an infalling or
outflowing wind.

\subsection{Extragalactic Ionizing Continuum Supplied by FR\,1s}

If the ionizing continuum radiation extrapolated from the HST/COS
spectra of M\,87 and NGC\,4696 is typical of FR\,1 radio galaxies,
then the contribution of this class of AGN to the extragalactic UV
background (UVB) in the local universe may be substantial since this
AGN class is so numerous \citep[$\sim$3\% of all bright cluster
ellipticals have $\log P(\rm W\,Hz^{-1})\geq22.0$;][]{LedlowOwen96}.

The beaming geometry in FR\,1s also must be taken into account in
estimating the amount of UVB contributed by these objects.  For
example, there is some evidence from BL\,Lac population studies
\citep[e.g.,][]{Perlman93, UrryPadovani95, Nieppola06, Padovani07,
Meyer11, Giommi13} that the amount of Doppler boosting (i.e, the
relativistic $\Gamma$) varies with frequency of emitted radiation.
This is also known as the re-collimating or accelerating jet model
\citep{Ghisellini89, UrryPadovani95, BoettcherDermer01} for blazars.
X-ray-selected BL\,Lacs (most of which are now termed High-energy
peaked BL\,Lacs or HBLs) often lack the optical non-thermal BL\,Lac
spectrum, instead showing either a pure stellar spectrum or a mixture
of an old stellar population and a non-thermal continuum.  This
property and their large space density means that the X-ray beam must
be much broader than the optical and radio emitting beams, presumably
coming from particles lower in $\Gamma$.  This hypothesis is supported
by the optical polarization studies conducted by \citet{Jannuzi94}
which showed that HBLs possess only modestly variable polarization
amounts and relatively constant position angles of polarization as if
we are viewing HBL optical beams from well off their axes.  A gradient
of broader to narrower emission beams (lower to higher $\Gamma$s) with
increasing wavelength predicts that the UV continuum beam is somewhat
narrower than the $\sim25-30$\arcdeg\ for the X-ray emission
\citep[e.g.,][]{Perlman93}. Due to this somewhat rudimentary
understanding of the beaming geometry, here we adopt a simplistic model
with a beamed ionizing flux emanating from two cones with half-angles
of 20\arcdeg\ and an unbeamed flux given by our FR\,1 observations emitted
over the rest of the sphere.

Since the extrapolated Lyman continuum flux of Hydra\,A it is almost
two orders of magnitude more luminous than that of M\,87, we assume
that some significant beaming is present in our direction towards this
source and we do not use its ionizing flux as typical of unbeamed
FR\,1s.  Using just M\,87 and NGC\,4696 as typical for the unbeamed
FR\,1 population, the ionizing continuum luminosity in these objects
is $\sim10^{40}$ ergs s$^{-1}$.  If only a small fraction of luminous
early-type galaxies are FR\,1s, then the ionizing radiation
contributed by this class is also quite small. But the evidence is
otherwise; e.g., \citet{LedlowOwen96} find that 3\% of all bright
cluster ellipticals have radio luminosities with $\log P_{1.4~GHz}
(\rm W\,Hz^{-1})\geq22$ while \citet{LinMohr07} find a larger fraction
of 5\% with $\log P_{1.4~GHz} (\rm W\,Hz^{-1}) \geq 23$ in X-ray
emitting clusters.  At X-ray wavelengths \citet{Martini06} found that
$\sim2$\% of all bright cluster ellipticals emit X-rays at
$P_x\geq10^{42}$ ergs s$^{-1}$. In a combined X-ray/radio study of
cluster AGN, \citet{Hart09} found 6\% of bright cluster ellipticals
are radio sources at $\log P_{1.4~GHz} (\rm W\,Hz^{-1})\geq 23.5$ and
1\% are X-ray sources at $P_x \geq 10^{42}$ ergs s$^{-1}$.
Extrapolation of these results to lower radio and X-ray power levels
is consistent with all bright cluster ellipticals being radio sources
at $\log P_{1.4~GHz} (\rm W\,Hz^{-1}) \geq 21.4$ and X-ray sources at
$P_x \geq 10^{40}$ erg s$^{-1}$.  Indeed, a deep {\it Chandra}
observation of the central region of the Perseus Cluster
\citep{Santra07} has detected weak ($10^{40-41}$ ergs s$^{-1}$)
non-thermal X-ray point sources from {\em all thirteen} early-type
galaxies brighter than $0.2\,L^*$ in this region.  In the
\citet{Santra07} X-ray imaging, there is no obvious trend in X-ray
luminosity with optical luminosity, suggesting that the AGN emission
in these X-ray point sources is similar in luminosity for all bright
cluster ellipticals.  Together, all these studies suggest that
virtually all bright early-type cluster galaxies are weak FR\,1s and
could be emitting Lyman continuum radiation at levels similar to
M\,87. Despite consistent results from the several studies cited
above, this important assumption must be confirmed in order to
validate our conclusions concerning the FR\,1 contribution to the UVB.

The CfA galaxy luminosity function of elliptical galaxies from
\citet{Marzke94} finds that the number density of $L>0.2L*$
ellipticals is $4\times10^{-3}$ Mpc$^{-3}$ (we ignore S0 galaxies in
this estimate since they appear not to harbor FR\,1s in general).
Following the \citet{Santra07} result as well as the extrapolations
from higher power levels made in the other studies quoted above, we
assume that all bright ellipticals emit Lyman continuum radiation at a
level similar to M\,87 and NGC\,4696. From this assumption we derive
an estimate of the unbeamed UV emission from FR\,1s of
$\sim4\times10^{37}$ ergs s$^{-1}$ Mpc$^{-3}$. This amount is
$\sim7$\% of the total ionizing background at $z\approx0$
\citep[assuming the total UV background at $z\sim0$ from][]{Haardt12}.

If the UV luminosity of $10^{40}$ ergs s$^{-1}$ is the unbeamed
continuum then beaming of these sources can add significantly to this
total, perhaps as much as an additional 7\% at $z\approx0$ if the
standard beaming model of \citet{UrryPadovani95} is assumed (half
opening angle of $\approx20$\arcdeg; see above). This standard model
assumption is equivalent to assuming that 1\% of all FR\,1s are seen
as beamed sources at any one location and that their mean observed
Doppler boosting factor is 20 in the UV. There are a very few (roughly
one in $10^6$) FR\,1s whose beamed non-thermal luminosities are at the
luminosities of $L_x>10^{44}$ ergs s$^{-1}$ \citep{Morris91} typical
of HBL type BL\,Lac objects, but these contribute negligibly to the
UVB total.

Overall the FR\,1 AGN class could contribute an amount to the ionizing
background of up to 1/7th (10-15\% of the total UVB) of the amount
contributed by QSOs and Seyferts locally.  If the above assumptions
about beaming and the FR\,1 source population are correct, luminous
ellipticals contribute non-negligibly to the amount of ionizing
background inferred from source population studies at low-$z$
\citep{Haardt01,Haardt12}. A sizeable contribution from FR\,1s could
account for the discrepancy between the most recent UV background
model of \citet{Haardt12} (which has minimal contribution from normal,
star forming galaxies) and the observed absorber statistics in the
\Lya\ forest \citep{Kollmeier14, Shull15}.  However, since most bright
ellipticals are found in rich clusters this putative additional UV
radiation could be very non-uniform, clumped around regions rich in
early-type galaxies. A more detailed look at FR\,1 radio galaxy
beaming geometries and number density statistics is required at assess
these possibilities in more detail.

Support for the idea that a large contribution to the UVB could be
made by FR\,1s/BL\,Lacs comes from the all-sky survey of far-UV
sources made by the {\it Extreme Ultraviolet Explorer} (EUVE). This
continuum survey conducted at a wavelength closer to the Lyman limit
than any other, detected more BL\,Lacs and narrow-line Seyferts than
any other AGN class \citep{Marshall95}.  These two AGN classes share
an unusual SED with a very steep soft X-ray spectrum as well as the
absence of broad permitted lines in their optical and UV spectra.  The
three FR1\,s studied here share the absence of broad emission lines
with these more luminous, EUV-bright AGN classes.

\section {Conclusions \& Summary}

We present far-UV spectroscopic observations of three FR\,1 radio
galaxies (M\,87, NGC\,4696, \& Hydra\,A) and two
previously-unpublished BL\,Lac objects (1ES\,1028$+$511 and
PMN\,J1103$-$2329).  All three FR\,1 galaxies show prominent \Lya\
emission at very close to the systemic redshifts and a weak FUV
continuum.

{\bf M\,87} is the brightest and most interesting of the objects
observed.  Though the COS/G130M spectral range is relatively short, we
measure a power law index of $\alpha_\lambda=0.87\pm0.06$ and
extrapolate the flux at the Lyman limit of $F_{912}=(1.78\pm0.04)
\times10^{-15}$~\flux\ at epoch 2014.

The strong \Lya\ emission of M\,87 is flanked by red and blue wings
separated from the main peak by $\sim1000-2000$~\kms.  Weak \NV, \CII,
and \SiIV\ emission lines are also seen at the redshift of M\,87 and
we confirm the very weak \FeXXI\ feature published by
\citet{AndersonSunyaev16} at a $5\sigma$ level.  

With the limited spatial information available with HST/COS, we find
that the emitting regions of both line and continuum are intermediate
between that of a point-source and a filled COS aperture implying that
both extend over regions tens of parsecs wide.  Furthermore, the
continuum may be slightly more extended than the line emission and the red
and blue wings of the \Lya\ emission are extended along the jet axis
consistent with being due to outflowing material from the nucleus.
The most puzzling feature of the M~87 FUV data is the possible very 
short ($\sim5$ day) \Lya\ line variability seen in the FOS data which is
inconsistent both with the spatial extent of the \Lya\ emission seen in
the COS aperture and the observed, rather narrow line widths of \Lya. 
\citet{Sankrit99} raise a concern that a slight displacement of the FOS 
aperture could have falsely created the apparent \Lya\ variability.
In light of the other evidence, the short-term \Lya\ variability
appears to be spurious although this can be confirmed or refuted by
future COS and STIS FUV spectroscopy (approved HST Cycle 24 program
14277).  It is expected that future long-slit STIS spectra also can
trace the extents of the three components of \Lya\ and the FUV
continuum to provide a viable physical picture of the M\,87 nucleus.

{\bf NGC\,4696} shows an asymmetric \Lya\ profile and a very weak UV
continuum.  We model the emission as a strong central component and
weak red wing overlaid on at least one strong \HI\ absorption system
at the systemic redshift.  We confirm the presence of weak \FeXXI\
emission in this object as well.

{\bf Hydra\,A} shows a simple broad-plus-narrow \Lya\ emission profile
and a moderate FUV continuum.  Metal-ion absorption lines in
\SiII$+$\OI\ and \CII\ show the presence of neutral, metal-enriched
gas associated with Hydra\,A, but it is not known if this gas is
circumnuclear or associated with the host galaxy on a broader scale.
The \Lya\ emission profile is well-fit without intrinsic \HI\
absorption.  Despite the appearance of an edge-on dust disk in optical
and FUV images of Hydra\,A \citep{Tremblay15}, the observations of
strong continuum and \Lya\ line emission--as well as the lack of
narrow \HI\ absorption seen in the other two FR\,1s--leads us to
conclude that the dust disk misses the nuclear region itself.

The BL\,Lac objects {\bf 1ES\,1028$+$511} and {\bf PMN\,J1103$-$2329}
show low-significance ($3-6\sigma$) \Lya\ emission features consistent
with other BL\,Lac objects from the literature (e.g., Paper~1; Fang
\etal\ 2014).  The \Lya\ luminosities of these objects are comparable
to that of the FR\,1s discussed here and in Paper~1.  While it is
generally accepted that FR\,1s are the parent population for BL\,Lac
objects (and in fact optical and X-ray imaging of BL\,Lacs reveal that
they reside in giant elliptical galaxies in clusters, as do FR\, 1s
\citep [see e.g.,][]{Wurtz96, Donato04, Sambruna07}, the emission
mechanisms, source locations and geometries for the beamed and
unbeamed ionizing continua and \Lya\ emission lines are still unclear.
As the nearest unobscured FR\,1 (Centaurus~A is a bit closer but has a
nucleus obscured by dust), M~87 provides the best opportunity to
observe the nuclear structures in ``radio-mode'' AGN and solve the
puzzle of their apparent very low accretion rates. New high spatial
and spectral resolution observations in the FUV with COS and STIS will
complement upcoming efforts at X-ray wavelengths and at mm wavelengths
with VLBI to determine the size and geometry of the line and
continuum-emitting regions in FR\,1s.

The observed \Lya\ luminosity in M\,87 and NGC\,4696 is within a
factor of a few to what is predicted by ``Case B'' recombination
illuminated by the non-thermal continuum in these sources extrapolated
to the Lyman edge.  Thus the circumnuclear gas could be either in an
ionization- or density-bounded regime.  The ionizing continuum of
Hydra\,A overpredicts the observed \Lya\ luminosity in this FR\,1 by a
factor of ten suggesting that the observed continuum has some
significant contribution from beamed radiation as well as the more
isotropic radiation seen in the two other FR\,1s. The extrapolated FUV
continuum flux of BL\,Lac objects overpredicts the observed \Lya\ line
flux by factors of $10^3-10^5$.  These results are consistent with all
of these sources possessing an unbeamed non-thermal continuum that is
the primary ionizing source for the gas that produces the observed
\Lya\ emission line.  While both the continuum and \Lya\ line are seen
to vary, the substantial distance (1-10 pc) we have inferred for the
\Lya\ emitting region does not allow the tight connection between
continuum and line variability that is seen for Seyfert galaxies
\citep[``reverberation mapping''; ][]{Peterson14}.  However, one
apparently successful narrow-line-region distance determination has
been obtained for NGC\,5548 using the [\ion{O}{3}] 5007\AA\ forbidden
line \citep{Peterson13}. A similar study may be possible for M~87. 

FR\,1 radio galaxies are very common in the modern era, comparable in space
density to Seyfert galaxies. Recent work at radio and X-ray wavelengths 
\citep*[e.g.,][]{Martini06, Santra07, LinMohr07, Hart09} make a
plausible case that virtually all bright ellipticals are AGN with
ionizing luminosities comparable to or greater than what we find in
M\,87 ($P_{UV}=10^{40}$ ergs s$^{-1}$). If it is assumed that all
$L\geq0.2L^*$ ellipticals possess unbeamed non-thermal continua at
this level, the contribution to the low-$z$ UVB could be substantial,
perhaps as much as $\sim10$\% of the extragalactic ionizing
background.

\medskip
\medskip
We acknowledge valuable input on this project from Mike Shull and Mike
Anderson.  This work was made possible by HST guest observing grant GO-13489
(J. Stocke, PI).

{\it Facility: HST (COS)}

\end{document}